\documentclass[conference]{IEEEtran}
\IEEEoverridecommandlockouts
\usepackage{cite}
\usepackage{amsmath,amssymb,amsfonts}
\usepackage{bm}
\usepackage{algorithmic}
\usepackage{graphicx}
\usepackage{textcomp}
\usepackage{xcolor}
\usepackage{epstopdf}
\def\BibTeX{{\rm B\kern-.05em{\sc i\kern-.025em b}\kern-.08em
    T\kern-.1667em\lower.7ex\hbox{E}\kern-.125emX}}
\begin{document}

\title{An Enhanced SCMA Detector Enabled by Deep Neural Network}

\author{
\IEEEauthorblockN{Chao~Lu, Wei~Xu, Hong~Shen, Hua~Zhang, and Xiaohu~You}
\IEEEauthorblockA{National Mobile Communications Research Laboratory, Southeast University, Nanjing 210096, China}
Email: \{220170709, wxu, shhseu, huazhang, xhyu\}@seu.edu.cn
}

\maketitle
\newtheorem{mylemma}{Lemma}
\newtheorem{mytheorem}{Theorem}
\newtheorem{mypro}{Proposition}
\begin{abstract}
In this paper, we propose a learning approach for sparse code multiple access (SCMA) signal detection by using a deep neural network via unfolding the procedure of message passing algorithm (MPA). The MPA can be converted to a sparsely connected neural network if we treat the weights as the parameters of a neural network. The neural network can be trained off-line and then deployed for online detection. By further refining the network weights corresponding to the edges of a factor graph, the proposed method achieves a better performance. Moreover, the deep neural network based detection is a computationally efficient since highly paralleled computations in the network are enabled in emerging Artificial Intelligence (AI) chips.
\end{abstract}


\section{Introduction}
The 5th generation (5G) wireless communication network aims at realizing expectations including extremely heavy connectivity, considerable high spectral efficiency, and ultra-low latency. Sparse code multiple access (SCMA) is a typical non-orthogonal multiple access (NOMA) strategy for 5G. It has been seen as a potential solution to address some of the critical requirements of 5G. Even though SCMA can deal with the troublesome issues in 5G, it faces two major challenges in terms of low-complexity detection  \cite{Wei2016} and efficient codebook design \cite{Alam2017}.

Message passing algorithm (MPA) is the most popular approach to implement multiuser detection with reduced complexity. Theoretically, it has been proved that if there is no loop in the factor graph of a, e.g., SCMA schema, the MPA gets exactly the marginal probabilities needed for detection \cite{Kschischang2001}, thus achieving the maximum likelihood (ML) boundary. While there inevitably exists loops, this algorithm is inherently suboptimal.

Recently, deep learning methods have been evidenced amazing progress in the fields of computer vision, speech recognition and natural language processing. Dramatic effects of deep learning have attracted a lot of attention. In traditional research fields of wireless communication, deep learning has been showing the promising ability to solve some specific problems \cite{Nachmani2016,Gruber2017,Sheam2017,Wen,Kim2018,Ye2018}. Some researches showed that deep learning methods are promising in performance enhancement of decoding \cite{Nachmani2016,Gruber2017}. Other researches treated a communication system as an end-to-end encoder and decoder network \cite{Sheam2017}. In addition, deep learning was proved successful in channel estimation by treating the channel matrix as a 2D image \cite{Wen}.

Besides the above studies, the investigation of applying deep learning approaches in wireless communication is still in its infancy. In this paper we consider the multiuser detection problem in the scenario of SCMA. By unfolding MPA and assigning weights to the edges of the factor graph, we construct a sparsely connected neural network. After training the neural network offline, we can achieve a better performance and the network can be deployed for online detection.

Note that a parallel work was presented in \cite{Kim2018} recently on the joint optimization of constellation mapping and detection for SCMA network. \cite{Kim2018} achieves great performance gain compared with the conventional method since the constellation mapping is also trained in the network which is crucial to the system performance. However if the constellation mapping is given, its neural network schema cannot outperform MPA. Or other, there will still be some distance between the neural network method proposed in \cite{Kim2018} and MPA if a much more better constellation mapping is given. Different from \cite{Kim2018}, we only study the detection problem, which means we fix the constellation mapping. From the results, our network can outperform the MPA at high signal noise ratio (SNR).

The remainder of the paper is organized as follows. In Section II, we describe the system model of SCMA structure. Then in Section III, we present the neural network based detection algorithm by transforming an MPA based detection procedure. In Section IV, the symbol error rate (SER) of the MPA and the neural network method is compared. Section V draws a conclusion of this work.

\section{SCMA Architecture}

We consider a SCMA network where there are \emph{J} independent users and \emph{K} orthogonal resource blocks. For a NOMA system, we generally have \emph{K} $ < $ \emph{J}. Each user has an \emph{M}-ary symbol set in which the symbols are assumed to be independent with equal probability. The ${\log _2}M$ symbols are one-to-one mapped to \emph{K} complex vectors by referring to a pre-designed table or codebook in other words. The \emph{K} vectors all have \emph{p} non-zero elements and (\emph{K-p}) zero elements. Fig. 1 exemplifies a system architecture of 6 users and 4 resources. The 4 resources are defined by the 4 square boxes. The blank box implies that the user has no symbol transmission assigned on this resource element. Mathematically, this process can be expressed in the following formula

\begin{figure}[htbp]
\centerline{\includegraphics[width=6.5cm,height=3.9cm]{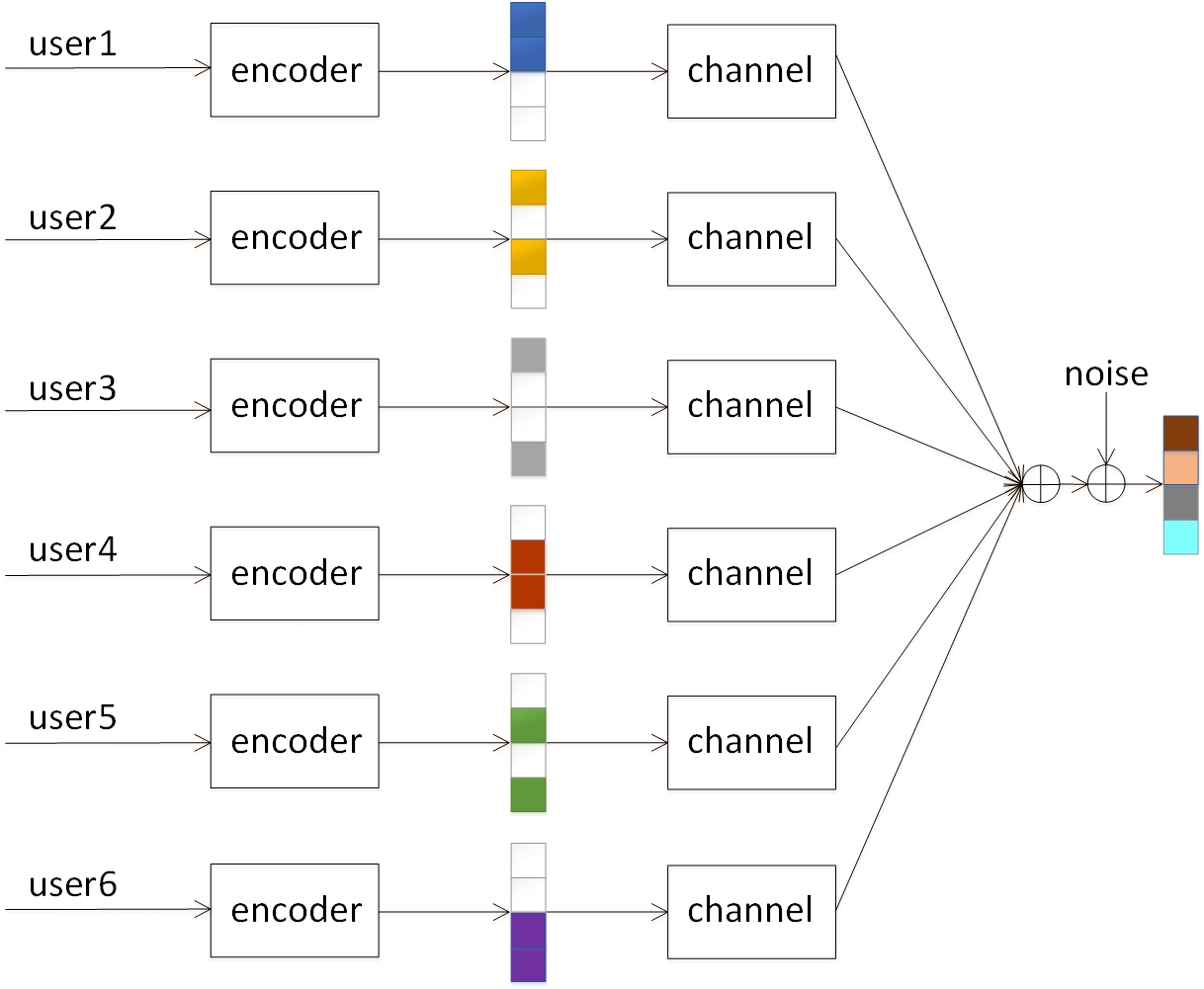}}
\caption{6-user-4-resource SCAM Architecture.}
\label{fig}
\end{figure}

\begin{equation}\label{eq:reveive-signal}
\mathbf{y} = \sum\limits_{j = 1}^J {{\mathop{\rm diag}\nolimits} ({\mathbf{h}_j}){\mathbf{x}_j}}  + \mathbf{n},
\end{equation}
where $\mathbf{y}={[{y_1},{y_2},...,{y_K}]^T}$ represents the received signal while ${\mathbf{x}_j}={[{x_{(j,1)}},{x_{(j,2)}},...,{x_{(j,K)}}]^T}$ represents the \emph{K}-ary complex constellation signals of the \emph{j}-th user at the transmitter. The channel of the \emph{j}-th user is denoted as ${\mathbf{h}_j}={[{h_{(j,1)}},{h_{(j,2)}},...,{h_{(j,K)}},]^T}$ and $h_{(j,k)}, k=1,2,...,K$ defines the channel of the \emph{j}-th user on the \emph{k}-th resource. As we can see from \eqref{eq:reveive-signal}, the signals of all \emph{J} users are added together on every resource block after passing the channel. Here, $\mathbf{n}$ represents the additive white Gaussian noise (AWGN) caused by the thermal noise of the amplifiers. The noise is a \emph{k}-ary vector and each element is subjected to a Gaussian distribution with mean of 0 and variance of ${\sigma ^2}$.

For efficient elaboration, we consider a typical SCMA network which has 6 users and 4 resources. As depicted in Fig. 2, the relationship between users and resources can be expressed by a factor graph. The edges between user nodes and resource nodes mean that these users have signals transmitted on these resource blocks. In Fig. 2, every resource has ${d_c}=3$ conflicting user signals. This means that the points of the constellation on every resource are sparse, which makes it possible to ensure the Euclidean distance of every two points large enough for reliable detection.

\begin{figure}[htbp]
\centerline{\includegraphics[width=7cm,height=3cm]{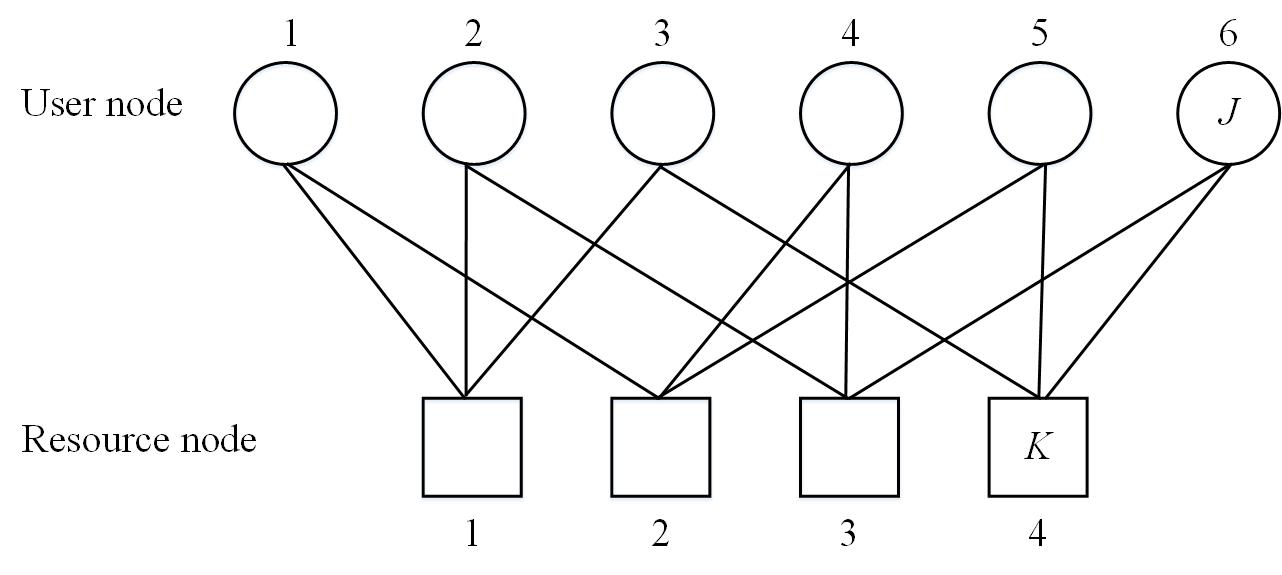}}
\caption{Typical factor graph.}
\label{fig}
\end{figure}

Alternatively, the factor graph can be represented by an indicator matrix $\mathbf{F}$. The factor graph in Fig. 2 equivalently corresponds to the indicator matrix:
\begin{equation}\label{eq:F-matrix}
{\bf{F}} = \left[ {\begin{array}{*{20}{c}}
{\begin{array}{*{20}{c}}
1&1&{\begin{array}{*{20}{c}}
1&0&0
\end{array}}&0
\end{array}}\\
{\begin{array}{*{20}{c}}
1&0&{\begin{array}{*{20}{c}}
0&1&1
\end{array}}&0
\end{array}}\\
{\begin{array}{*{20}{c}}
0&1&{\begin{array}{*{20}{c}}
0&1&0
\end{array}}&1
\end{array}}\\
{\begin{array}{*{20}{c}}
0&0&{\begin{array}{*{20}{c}}
1&0&1
\end{array}}&1
\end{array}}
\end{array}} \right].
\end{equation}

Using $\mathbf{F}$, we define two sets,

\begin{equation}\label{eq:Vk}
V(k){\rm{ = \{ }}j|{\left[ {\bf{F}} \right]_{k,j}} = 1{\rm{\} }},{\rm{ }}k = 1,{\rm{ }}2,{\rm{ }} \cdots ,{\rm{ }}K
\end{equation}
and
\begin{equation}\label{eq:Cj}
C(j){\rm{ = }}\left\{ {k|{{\left[ {\bf{F}} \right]}_{k,j}} = 1} \right\},{\rm{ }}j = 1,{\rm{ }}2,{\rm{ }} \cdots ,{\rm{ }}J,
\end{equation}
where \emph{V(k)} represents the set of users that reuse the \emph{k}-th resource block while \emph{C(j)} represents the set of the resource blocks that the \emph{j}-th user occupies. \emph{V(k)} and \emph{C(j)} will be used for illustration in the following part.

\section{From MPA to Neural Network}

MPA calculates the marginal probabilities through iterations. In this work, we replace the iteration steps by neural network layers. Next, we will explain how MPA and our neural network work.

\subsection{Message Passing Detection Algorithm}
The detection of SCMA is conducted by maximizing the posterior probability

\begin{equation}\label{eq:post-prob-joint}
{\hat{\bf{X}}} = \arg \mathop {\max }\limits_{{\bf{X}} \in {{\bf{X}}^{J,K}}} p({\bf{X}}|{\bf{y}}),
\end{equation}
where ${\bf{X}} = [{{\bf{x}}_1},{{\bf{x}}_2},...,{{\bf{x}}_J}]$ and ${{\bf{x}}_j} \in {{\bf{X}}_j},j = 1,2,...,J$. ${{\bf{X}}_j}$ represents the set of the \emph{j}-th user’s constellation signals, while ${{{\bf{X}}^{J,K}}}$ is all the combinations of possible constellation signals from different users. By calculating the marginal probability in \eqref{eq:post-prob-joint}, the decision formula for each user can be written as

\begin{equation}\label{eq:post-prob-single}
{{\hat{\bf{x}}}_j} = \arg \mathop {\max }\limits_{{{\bf{x}}_j} \in {{\bf{X}}_j}} \sum\limits_{{\bf{X}} \in {{\bf{X}}^{J,K}},{{\bf{x}}_j} = \left[ {\bf{X}} \right]_j } {p({\bf{X}}|{\bf{y}})} ,
\end{equation}
where ${{\hat{\bf{x}}}_j}$ is the estimation obtained by using, e.g., MPA for solving \eqref{eq:post-prob-single}.

\begin{figure*}[htbp]
\centerline{\includegraphics[width=15cm,height=6.3cm]{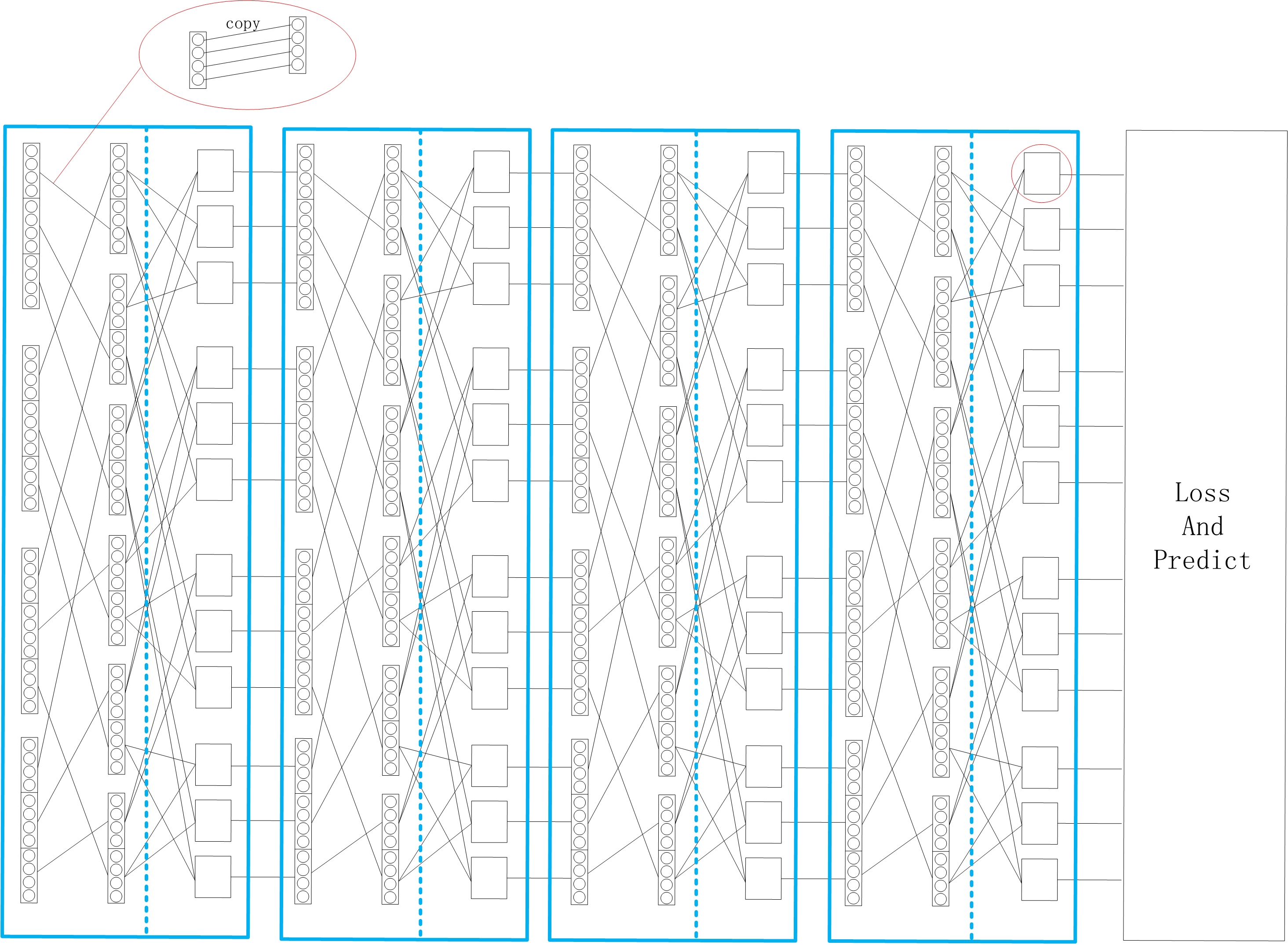}}
\caption{Structure of neural network, every blue square represents a block and the block in red circle is named as pooling-concat block.}
\label{fig}
\end{figure*}

In MPA, the posterior probability is calculated iteratively. Two types of messages are calculated and then exchanged in the factor graph. The message transmitted from the \emph{k}-th resource to the \emph{j}-th user is defined by  $I_{{R_k} \to {u_j}}^t$. The message transmitted from the \emph{j}-th user to the \emph{k}-th resource is defined by $Q_{{u_j} \to {R_k}}^t$. The superscript \emph{t} represents the iteration index. The two types of messages are updated according to the following equations.

\begin{equation}\label{eq:I-MPA}
\begin{aligned}
I_{{R_k} \to {u_j}}^t({{\bf{x}}_j}) &= \sum\limits_{{{\bf{X}}^{V(k)\backslash j}}} {p({y_k}{\rm{|}}{{\bf{x}}_{{j_1}}},{j_1} \in V(k))} \\
& \quad \times \mathop \prod \limits_{{j_2} \in V(k)\backslash j} Q_{{u_{{j_2}}} \to {R_k}}^{t - 1}({{\bf{x}}_{{j_2}}})\\
& = \sum\limits_{{{\bf{X}}^{V(k)\backslash j}}} {\frac{1}{{\sqrt {2\pi {\sigma ^2}} }}} \exp \left( { - \frac{1}{{2{\sigma ^2}}}{A_k}} \right)\\
& \quad \times \mathop \prod \limits_{{j_2} \in V(k)\backslash j} Q_{{u_{{j_2}}} \to {R_k}}^{t - 1}({{\bf{x}}_{{j_2}}}),
\end{aligned}
\end{equation}

\begin{equation}\label{eq:Ak}
{A_k}{\rm{ = }}{\left\| {{y_k} - \sum\limits_{{j_1} \in V(k)} {{h_{(k,{j_1})}}{x_{(k,{j_1})}}} } \right\|^2},
\end{equation}

\begin{equation}\label{eq:Q-MPA}
Q_{{u_j} \to {R_k}}^t({{\bf{x}}_j}) = p({{\bf{x}}_j})\mathop \prod \limits_{k' \in C(j)\backslash k} I_{{R_{k'}} \to {u_j}}^{t - 1}({{\bf{x}}_j}),
\end{equation}
where $C(j)\backslash k$ means the set of $C(j)$ after removing component \emph{k} and $V(k)\backslash j$ means the set of $V(k)$ after removing component \emph{j}. ${{\bf{X}}^{V(k)\backslash j}}$ represents all possible combinations of symbols sent by all other users in set $V(k)$ expect for the \emph{j}-th user.

Multiplication can be replaced by addition in logarithmic domain. Usually, equations \eqref{eq:I-MPA} and \eqref{eq:Q-MPA} can be rewritten as

\begin{equation}\label{eq:LI-MPA}
\begin{aligned}
LI_{{R_k} \to {u_j}}^t({{\bf{x}}_j}) &= \ln \biggl( \sum\limits_{{{\bf{X}}^{V(k)\backslash j}}} \exp \Bigl( - \frac{1}{{2{\sigma ^2}}}{A_k} + \\
 & \quad \sum\limits_{{j_2} \in V(k)\backslash j} {LQ_{{u_{{j_2}}} \to {R_k}}^{t-1}({{\bf{x}}_{{j_2}}})} \Bigr) \biggr) + \beta \\
 & \approx \mathop {max}\limits_{{x_j}} \biggl( { - \frac{1}{{2{\sigma ^2}}}{A_k}} +  \\
 & \quad {\sum\limits_{{j_2} \in V(k)\backslash j} {LQ_{{u_{{j_2}}} \to {R_k}}^{t-1}({{\bf{x}}_{{j_2}}})} } \biggr) + \beta,
\end{aligned}
\end{equation}

\begin{equation}\label{eq:LQ-MPA}
LQ_{{u_j} \to {R_k}}^t({{\bf{x}}_j}) = \ln (p({{\bf{x}}_j})) + \sum\limits_{k' \in C(j)\backslash k} {LI_{{R_{k'}} \to {u_j}}^{t - 1}({{\bf{x}}_j})},
\end{equation}
where $\beta  = \ln (1/\sqrt {2\pi {\sigma ^2}} )$ is a constant value.

\begin{figure}[htbp]
\centerline{\includegraphics[width=7cm,height=5.5cm]{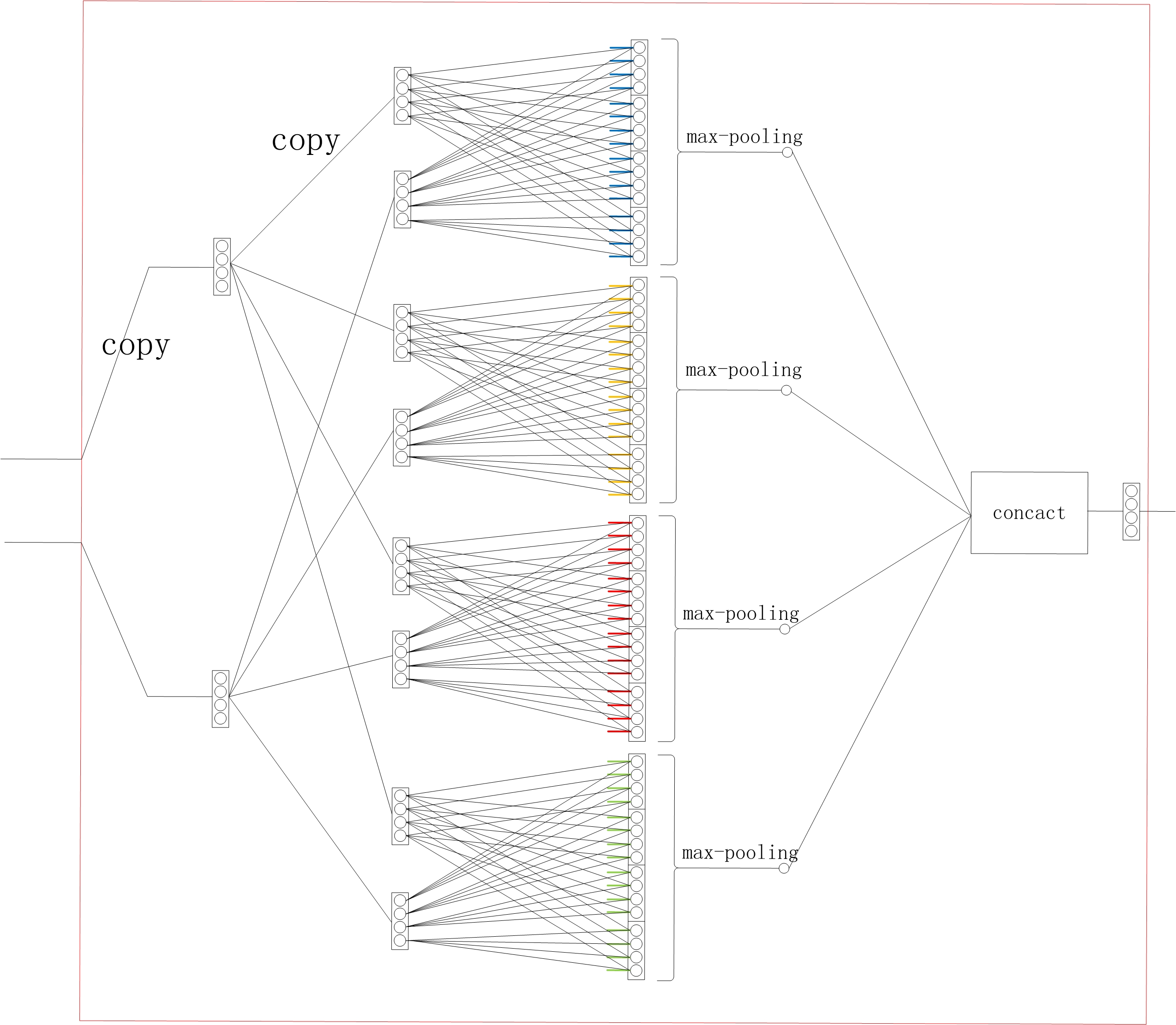}}
\caption{Pooling-concat block, small bold color lines are the inputs of the network.}
\label{fig}
\end{figure}

\subsection{Neural Network Architecture}
Regarding for computations of MPA, it is possible to replace manipulations with the propagation of a neural network. We consider a deep neural network of 4 blocks and each block consists of two layers divided by imaginary lines, as depicted in Fig. 3. Let us focus on the major computation steps in \eqref{eq:LI-MPA} and \eqref{eq:LQ-MPA} of MPA. Each iterative computations of \eqref{eq:LI-MPA} and \eqref{eq:LQ-MPA} can be unfolded as the propagation between two layers in a block. The pooling-concat block shown in Fig. 4. is calculated according to

\begin{equation}\label{eq:LI-NN}
\begin{aligned}
LI_{{R_k} \to {u_j}}^{2(l + 1)}({{\bf{x}}_j}) &= \mathop {\max }\limits_{{{\bf{X}}^{V(k)\backslash j}}} \biggl(  - \frac{{{c_{2(l + 1),{j_1},j,k}}}}{{2{\sigma ^2}}}{A_k} + \\
& \quad \sum\limits_{{j_2} \in V(k)\backslash j} {{w_{2(l + 1),{j_2},j,k}}LQ_{{u_{{j_2}}} \to {R_k}}^{2l + 1}({{\bf{x}}_{{j_2}}})} \biggr) \\
& \quad + {a_{j,k}}\beta.
\end{aligned}
\end{equation}

In Fig. 2, ${d_c}$ is equal to 3, so $V(k)\backslash j$ has 2 elements. As we can see in Fig. 4, the pooling-concat block has 2 inputs which denote $LI_{{R_k} \to {u_j}}^{2l+1}$. When \emph{M} is set to 4, each user will have 4 symbols, thus each input of the pooling-concat block has 4 nodes. The small bold color lines linked to the neurons before max-pooling operation are the inputs of calculated ${A_k}$. As for $\beta$ in \eqref{eq:LI-NN}, it exists as an offset in a neuron. The neurons assign different weights to different inputs and finally output their summation. The outputs of the neurons are followed by a maximum operation and the results will be merged to a 4-nodes chunk since ${{\bf{x}}_j}$ has 4 values.

The first layer in the block is calculated according to
\begin{equation}\label{eq:LQ-NN}
\begin{aligned}
LQ_{{u_j} \to {R_k}}^{2l{\rm{ + }}1}({{\bf{x}}_j}) &= {b_{2l{\rm{ + }}1,j,k}} \ln \left( {p({{\bf{x}}_j})} \right)+ \\
& \quad \sum\limits_{k' \in C(j)\backslash k} {{w_{2l{\rm{ + }}1,k',j,k}}LI_{{R_{k'}} \to {u_j}}^{2l}({{\bf{x}}_j})}.
\end{aligned}
\end{equation}

The front 4 chunks serve as placeholders to store the inputs from the previous block, usually $LQ_{{u_j} \to {R_k}}^{(2l+1)}$. The outputs of the placeholders are all set to $\ln \left( 1/M \right)$ if there are no inputs such as the first block. Similar to the pooling-concat block, the neurons of the back 6 chunks also output the weighted summation of their inputs.

In \eqref{eq:LI-NN} and \eqref{eq:LQ-NN}, it can be seen that we do not apply any activation functions which are often used to ensure the non-linear of the neuron output. To ensure the equivalence with MPA, we just let the linear combination of the inputs of neurons become the outputs.

As for the output layer, we follow the equation \eqref{eq:output} to calculate the output logits of the network, which is also applied in MPA.

\begin{equation}\label{eq:output}
L{Q_{{u_j}}}({\bf{x}}_j) = \ln \left( {p({{\bf{x}}_j})} \right) + \sum\limits_{k \in C(j)} {LI_{{R_k} \to {u_j}}^L({{\bf{x}}_j})}
\end{equation}

Finally, we can decide the estimate symbol by

\begin{equation}\label{eq:predict}
{\bf{x}}{'_j} = \arg \mathop {\max }\limits_{{{\bf{x}}_j}} L{Q_{{u_j}}}({{\bf{x}}_j}){\rm{,  }} \quad j = 1,{\rm{ }}2, \cdots ,{\rm{ }}J .
\end{equation}

After designing the structure of the neural network, we construct the following loss function as our optimization objective for network training:

\begin{equation}\label{eq:loss}
Loss = {\rm{E}} \left\{ \sum\limits_j {{{I}}({{\bf{x}}_j},{\bf{x}}_j')} \log (\frac{{\exp (L{Q_{{u_j}}}({{\bf{x}}_j}))}}{{\sum\limits_{{{\bf{x}}_{j'}}} {\exp (L{Q_{{u_j}}}({{\bf{x}}_{j'}}))} }})\right\},
\end{equation}
where the function ${{I}}({{\bf{x}}_j},{\bf{x}}_j')$ is an indicator function and it has the following form

\begin{equation}\label{eq:indicator}
{I}({{\mathbf{x}}_j},{\mathbf{x}}_j') = \left\{ {\begin{array}{*{20}{c}}
  {1, \quad {{\mathbf{x}}_j} = {\mathbf{x}}_j'} \\
  {0, \quad {{\mathbf{x}}_j} \ne {\mathbf{x}}_j'}
\end{array}} \right.,
\end{equation}
and ${\rm{E}}\{ \cdot \} $ represents mathematical expectation.

This loss function can be regarded as a softmax cross entropy.  More specifically, the 4 neurons' outputs are normalized by a softmax function separately. Then, the normalized results are used to calculate the cross entropy. Since it is impossible to calculate the mathematical expectation during training, we can replace the function ${\rm{E}}\{ \cdot \} $ by batch average mean function.

\section{Numerical Experiments}
In this section, we describe all the key details of our experimental tests. All of our the simulation uses the codebook proposed in \cite{Qualcomm2016} and its constellation graph is given in Fig. 5. In our experiment, the SCAM network has 6 users and every user has a set of 4 symbols. This corresponds to the output layer with 24 neurons.

\begin{figure}[htbp]
\centerline{\includegraphics[width=7cm,height=6cm]{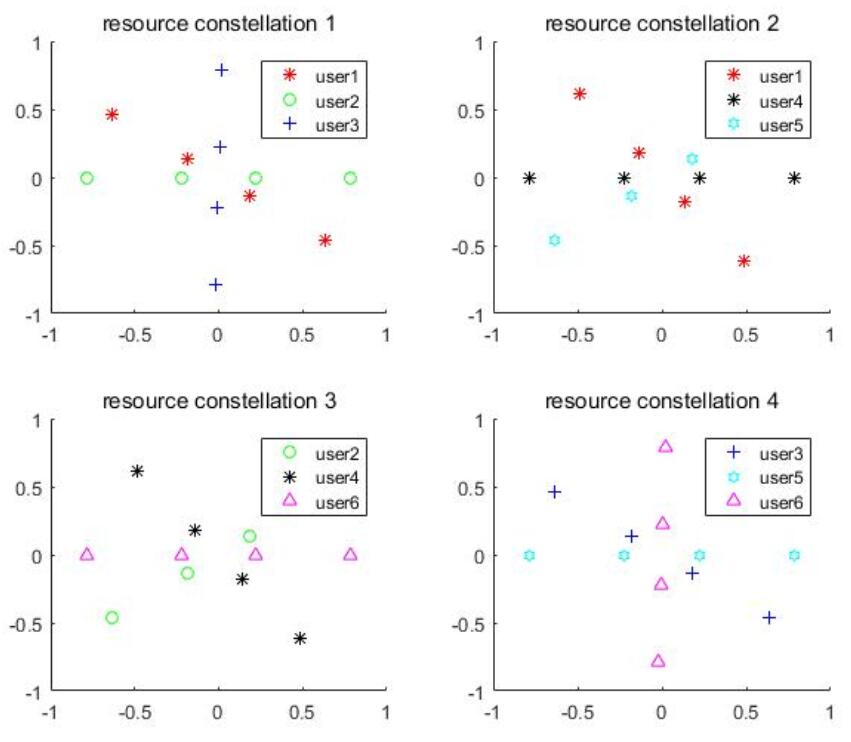}}
\caption{ Constellations on 4 resources.}
\label{fig}
\end{figure}
\subsection{Neural Network Training}
We build our neural network by the TensorFlow \cite{Abadi2015} framework. TensorFlow is an open source machine learning library provided by Google. Because of its encapsulation, ease of use, and free for academic research, we choose this framework to implement our neural network.

In order to reduce the training time and achieve a better result, it is important to give the neural network a good initialization. We initialize all the parameters with the same non-zero value such as all-in-one. We set the  initial learning rate value to 0.001, and then select the Adam \cite{Kingma2014} optimizer to optimize all network parameters. It is much more suitable to use such an advanced gradient optimizer because this optimizer can adjust learning rate to learning steps while a raw stochastic gradient optimizer cannot. We trained two neural networks with 2 blocks and 4 blocks, respectively. Both the networks are trained under the AWGN channel at a predetermined SNR. But after training, these networks effectively work at an arbitrary SNR.

\subsection{Dataset}
Here, we consider the SCMA setup where \emph{J} is 6, \emph{K} is 4 and \emph{M} is 4. The 6 users have 4096 signal combinations in total. We generate the training data dynamically which means the noises are all different at each step. The mini batch feed to the network is fixed to 4096 (the total combinations) and the training data is generated at a fixed SNR (16dB). We generate the test data in the same way, but at 3dB intervals.

\begin{figure}[htbp]
\centerline{\includegraphics[width=7cm,height=6cm]{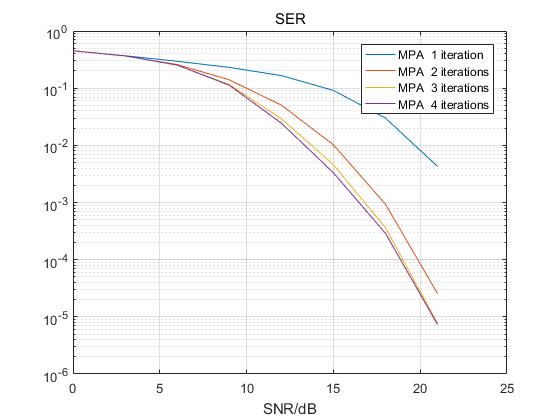}}
\caption{SER of MPA with different iterations at AWGN.}
\label{fig}
\end{figure}
\begin{figure}[htbp]
\centerline{\includegraphics[width=7cm,height=6cm]{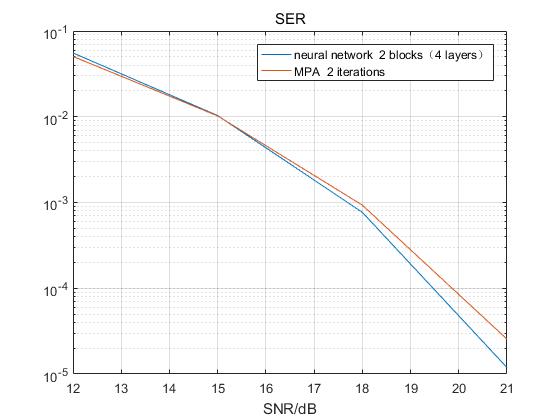}}
\caption{SER of MPA and neural network at AWGN.}
\label{fig}
\end{figure}
\begin{figure}[htbp]
\centerline{\includegraphics[width=7cm,height=6cm]{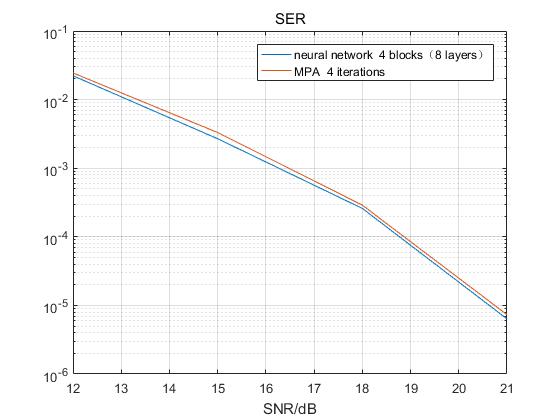}}
\caption{SER of MPA and neural network at AWGN.}
\label{fig}
\end{figure}

\subsection{Results}

In Fig. 6, we compare the SER performance of traditional MPA with various iteration numbers. As is shown in Fig. 6, with the growth of iteration numbers, the performance gets better. On the other hand, the performance gain becomes less with the increase of iteration numbers. When the iteration numbers come to 4, the algorithm performance almost converges.

For comparing, we give the SER performance of MPA and neural network method in Fig. 7 and Fig. 8. The neural network blocks are set to 2(4) when the MPA iterations  are set to 2(4) for a fair comparison. Obviously, the neural network method show the potential to achieve a better performance at high SNR. From Fig. 7 we can see that the curves of the two method intersect together at 15 dB. After that point, the curve of the neural network method shows some improvement. Similarly, this phenomenon also appears in Fig. 8. In other words, the neural network can find a better solution of this problem after training.

\section{Conclusion}
In this article, we propose a neural network architecture for SCMA detection. The result of this method converges faslty when the number of iterations is up to four. We show how to unfold the MPA to a neural network in detail. By adding different weights to the edges of the factor graph, the MPA is represented as a sparsely connected neural network. This neural network outperforms the traditional MPA especially at high SNR. Moreover, the neural network based method can be accelerated by GPU or AI processor, which also makes sense.

\section*{Acknowledgment}

This work was support by the National Natural Science Foundation of China under Grants 61471114, 61521061, 61501110, 61601115, U1534208, Six talent peaks project in Jiangsu Province under GDZB-005, the Natural Science Foundation of Jiangsu Province under Grant BK20150635, and the Fundamental Research Funds for the Central Universities under Grant 2242014K40037.

\end{document}